# Analysis of *Pseudomonas aeruginosa* Virulence in *Dictyostelium discoideum* Strains

**Asher Parvu and Katherine Ortiz**

*Department of Biological Sciences and Physical Sciences, University of Arkansas Fort Smith, 5210 Grand Avenue, Fort Smith, AR 72913-3649*
*Project was undertaken in a Biomedical Research course offered at the University of Arkansas Fort Smith.*

## Abstract
___________________________________________________________________

*Dictyostelium discoideum* is an ideal organism to test the pathogenicity of *Pseudomonas aeruginosa*, since its eukaryote nature allows for comparisons to be made to human health. Additionally, *D. discoideum* naturally feeds on bacteria and has a simplistic genome which improves result interpretation. *P. aeruginosa* is a free living bacteria that is recognized for its ability to cause nosocomial, hospital contracted, infections in immunocompromised patients. This experiment attempts to understand how various strains of *D. discoideum* may be more susceptible or resistant to infections caused by *P. aeruginosa*, as a result of environmentally induced adaptive mechanisms. Wild-type strains of *D. discoideum* will be isolated from Massard Prairie soil and compared to axenic-type strains that have been cultured free from association with other microbiota. The results of this experiment may support studies done on conserved virulence pathways in bacteria and give insight into virulence of *P. aeruginosa* in humans. There is evidence that *Dictyostelium* has capabilities that allow for protection against bacteria similar to *P. aeruginosa*.[16] Therefore, it is expected that the wild-type *D. discoideum* will have adaptive mechanisms that which will cause resistance to *P. aeruginosa*, unlike axenic-type strains that will have limited resistivity.

# Introduction
___

The soil amoeba, *Dictyostelium discoideum*, has been recognized globally for being a model organism because of its genetic similarities to many eukaryotic organisms. The simplicity of this organism has been useful in studies that relate to genetics, chemical, and cellular process. There have been numerous experiments that have studied the virulent effects *Pseudomonas aeruginosa* on axenic-type *D. discoideum* in attempts of identifying conserved virulent pathways that contribute to infection. It has been determined that there are two main pathways in mutant bacteria which are used to infect amoebae, quorum sensing mediated virulence and type III secretion of virulence factors.[1,11] Quorum sensing virulence refers to the bacteria's ability to release chemical signals as the concentration of bacteria grows. This physiological effect has been shown to have many implications in bacterial functions. Type III secretion systems have also been found in many pathogenic bacteria and have been identified in the direct injection of cytotoxins into host cells, this leads to infection and antibiotic resistance.[1,12] *P. aeruginosa* is an opportunistic bacteria that is known to cause nosocomial infections that can lead to severe illness. It has been determined that the virulence of *P. aeruginosa* is controlled by two quorum sensing pathways, *las* and *rhl*, which leads to the secretion of virulence factors.[13] It is capable of infecting various host systems of varying complexities, therefore *D. discoideum* is a good alternative to other more expensive and complicated organisms.[2] The experiment will isolate *D. discoideum* soil samples obtained from undisrupted areas of Massard Prairie and disrupted areas of Massard Prairie found in Ben Geren Park. These locations should have different soil contents that is since one area is exposed to a multitude of chemicals and the other is not. The proposed experiment would analyze the virulence of *P. aeruginosa* towards wild type *D. discoideum* obtained from two different soil samples and against axenic-type *D. discoideum*. Utilizing two different wild-type *D. discoideum* strains, obtained from the Massard Prairie soil samples, will insure that any virulent resistance and susceptibility can be analyzed based on the native environment of the organism. One wild-type strain will be obtained from unadulterated Massard Prairie soil, which will ensure that that normal wild-type processes are developed in the organism according to its natural environment. The other wild-type strain of *D. discoideum* will be isolated from soil obtained from Ben Geren Park, since it may contain an assortment of chemicals. It is expected that microorganisms and chemicals present in the soil will influence the amoeba's susceptibility to bacterial virulence. Isolated wild-type strains along with an axenic amoeba strain will be inoculated with *P. aeruginosa* to test virulence resistivity. *P. aeruginosa* may have implications in research that are focused on addressing areas of biomedical research focused on antibiotic resistance and adaptive immunity. Many studies utilize isolated axenic strains developed in laboratories, such as Northwestern University in Chicago, which ensures integrity of axenic nature. Axenic-type strains are not in association with other microbiota during cultivation, therefore, they may have less adaptive strategies developed. Many experiments rely on axenic-type strains of *D. discoideum* to gather simplified results that are not influenced by extrinsic factors. As a result, only a limited number of experiments have been performed to test the virulent resistance of non-axenic type *D. discoideum* to *P. aeruginosa*. The hypothesis is that wild-type *D. discoideum* will have developed adaptive capabilities, resulting in increased resistance to *P. aeruginosa* pathogenicity.

## Materials and Methods
________________________________________________________________

### Isolation of wild-type *D. discoideum* from soil samples

There are nine species of *Dictyostelium* along with countless other species of soil amoeba that have been identified, and likely many more that have yet to be discovered. In a soil sample it is common to obtain many species of *Dictyostelium* along with other organisms, such as mycobacteria, slime molds, pathogens. Wild-type strains *D. discoideum* will be obtained from soil samples taken at two locations of Massard Prairie in Fort Smith, Arkansas **FIG. 1**. Currently, the Massard Prairie is being restored so there may be remnants of prairie being treated with potential fertilizer and chemicals **FIG. 2**. Further collaboration will be needed with Ben Geren Parks management to identify whether amoeba should be isolated from native or restored regions.

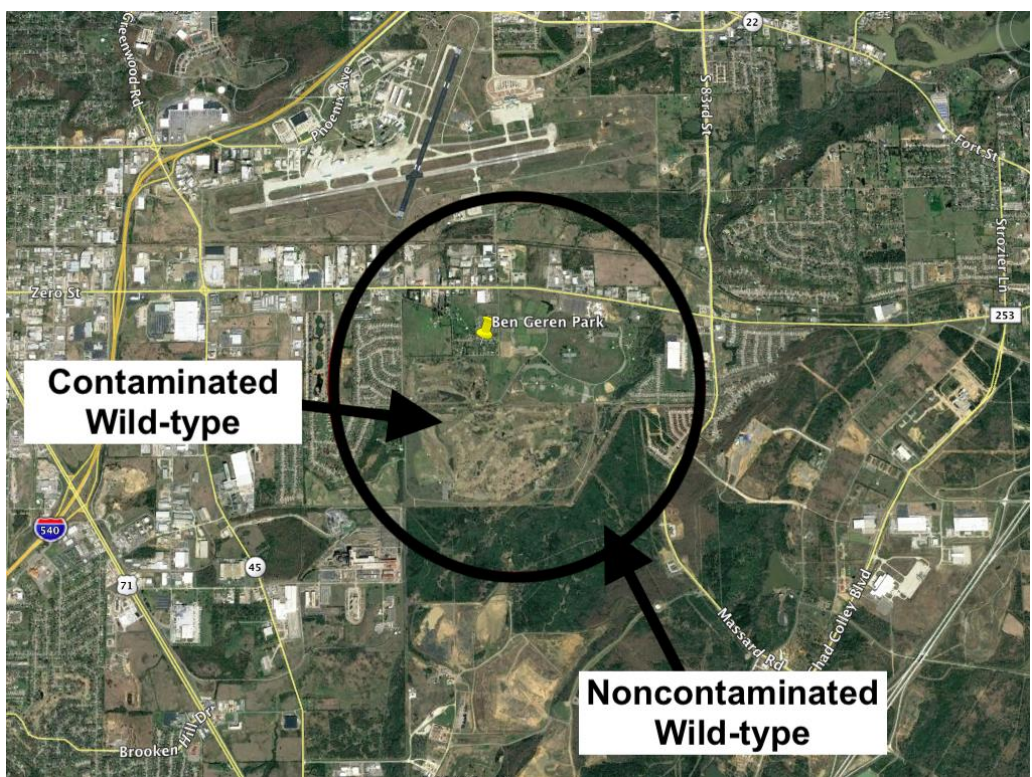

**FIG. 1.** Ben Geren Park located on the Massard Prairie[10]

One of the locations will be a remnant of the native untouched prairie; the other will obtained from Ben Geren Park, also a remnant of the prairie, that may continue to undergo toxic treatments of synthetic fertilizers, fungicides, insecticides, herbicides, and other chemicals.

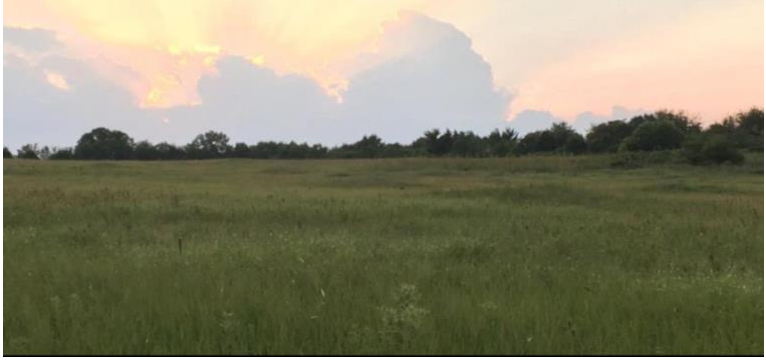
**FIG. 2.** Massard Prairie Restoration Project[10]

Wild-type *D. discoideum* grows best in area of high moisture, so samples of soil will be obtained from under decomposing logs, next to trees, or streams. When collecting soil samples, gloves should be worn and a clean spoon used to handle sample. Soil should be kept in a Ziploc bag and labeled (location, sample number). Once samples are obtained the begin experiment by making SM agar plates. SM agar plates are made by mixing: 10g glucose, 10g protease peptone, 1g $MgSO_4 \times 7H_2O$, 1.9g $KH_2PO_4$, 0.6g $K_2HPO_4$, 20g bacto agar, 900 mL distilled water. The mixture was autoclaved for 20 minutes for sterilization and 35 mL was poured into each plate. 1 liter of SM broth was made by mixing 10 g glucose, 10 g proteose peptone, 1 g yeast extract, 1 g $MgSO_4 \times 7H_2O$ (or 0.5 g $MgSO_4$), 1.9 g $KH_2PO_4$, 0.6 g $K_2HPO_4$. Titrations of KOH were used to adjust the pH between 6.0 - 6.4. In a 10 mL tube, make a 1:10 dilution for each soil sample by adding 1.0 gram of the sample to 9.0 mL of distilled water, and vortex for 5-10 seconds. Immediately after following the suspension of soil in water, make a second tenfold dilution (take another 10 mL tube) by pipetting 1.0 mL (debris should not be pipetted) of the 1:10 dilution into 9.0 mL of distilled water. This is 1:100 dilution. Vortex for 5-10 seconds. Make a final 1:1000 dilution by adding 1.0 mL of the 1:100 dilution to 9.0 mL of distilled water. Vortex for 5-10 seconds. Take 200 μL of the 1:1000 suspension in a 1.5 mL tube. To this tube add 300 μL of a non-pathogenic *E. coli* culture (grown in SM media). Vortex each sample tubes for 5-10 seconds. Pipette 500 μl sample/*E.coli* suspension and spread onto SM agar plate. Plate *D. discoideum* cells grown on SM media with *E.coli* to allow comparison of the isolated strains to a known strain. The plates should be incubated (with upside down) for 1 week at 22°C (low temperature incubator). Following the incubation period, plates should be inspected under stereomicroscope or inverted microscope for evidence of amoebae. Look for clearing zones in the agar or fruiting bodies. Various filamentous fungi will spread over the plate with time, so plates need to be checked frequently to find colonies before they are overgrown. If you see fruiting bodies or plaques, use a pipette tip to carefully harvest them. Suspend the harvest in 400 μL of the *E.coli* culture before plating the *E.coli* + isolate on a SM plate. The fruiting bodies from these plates should be harvested and re-plated in this same manner until there is no more evidence of other organisms competing or coexisting with the wild-type *D. discoideum*.[3]

**Initiation of axenic culture of *D. discoideum* from wild-type plate growth**

*D. discoideum* cells can easily be transferred from a plate to axenic growth in suspension. Separation of bacteria and *D. discoideum* occurs owing to the difference in weight between the two cell types. Use a sterile loop to remove amoebae from a single colony from either bacterial growth plate bacterial plate (use vegetative cells from the feeding front at the perimeter of the plaque, where cells are feeding and dividing) and suspend in 0.2 ml fresh bacterial overnight culture. Using a sterilized glass spatula, spread the bacteria-wild-type *D. discoideum* mixture onto an SM agar plate and incubate at 22°C. Harvest the cells when the plates have been cleared of bacteria but before fruiting bodies have formed (2-3 days) by flooding the plate with 3-4 ml of DB. Recover cells by gently scraping with a sterile spatula, transfer to a 50 ml conical plastic tube, fill with 40 mL DB, briefly vortex and centrifuge at 500*g* for 4 min at room temperature. Discard the cloudy supernatant containing the bacteria and resuspend the *D. discoideum* cells in 45 ml DB. Wash four more times to remove as many bacterial cells as possible. Resuspend the final pellet in 25 mL HL5 containing streptomycin and ampicillin and transfer to a 125 ml Erlenmeyer flask. The medium should only contain 0.03 grams in streptomycin per liter of nutrient medium, as its purpose is to prevent bacterial contamination of *Dictyostelium*. If the culture is kept open for prolonged periods of time the culture is has a higher likelihood to become contaminated. Adding a minimal amount of streptomycin will also prevent it from cause genetic mutations in *Dictyostelium* and changing morphology. If *D. discoideum* morphology is changed the results become more complex and it becomes difficult to decipher whether the results are permissible. Next, the Erlenmeyer flask should be incubated for 2-3 days at 22°C, shaking at 180 r.p.m. Transfer a small volume (1-2 ml) into a 250 mL flask containing 50 mL fresh HL5; at this point, the addition of antibiotics is optional. These cells can now be maintained axenically as described. To exclude bacteria, make sure that the supernatant is cloudy after the first centrifugation, indicative of a large number of bacteria remaining in suspension; if not, reduce the speed and/or the time of centrifugation. Cloudiness should diminish with each washing step.[4] Axenic-type *D. discoideum* cells may be requested from the Dicty Stock Center, a central repository for *Dictyostelium discoideum* strains, in order to circumvent the axenic culture protocol. In conducting this experiment, it is also advisable to plate all wild-type strains of *D. discoideum* with *E. coli* bacteria so these control groups can be used for comparisons.

**Inoculation of *P. aeruginosa* to wild-type and axenic-type strain *D. discoideum***

In this experiment virulence susceptibility of wild-type and axenic-type strains of *D. discoideum* to *P. aeruginosa* can be determined by inoculating plates and measuring the size of amoeba induced plaques. All cell cultures of *D. discoideum* cells strains need to be diluted equally to insure that concentration of plated amoeba is proportional to added bacteria. The protocol to accomplish this begins by identifying initial cell concentration of all cultured amoeba strains and diluting to a set concentration using a HL5 medium. If the initial concentration of cultured amoeba is $6x10^7$cells/mL, then it will be diluted by the addition of 5999 µL (5.999 mL) of HL5 medium. Next, 100 µL of the diluted culture (100 dicty cells) will be placed on the SM agar plate that contains 200 µL of *P. aeruginosa* bacterial culture. This protocol would be applied to both wild-types and the axenic-type strain of amoeba. When *D. discoideum* is grown on bacteria, the cells have a doubling time that is approximately 4 hours, while cells grown in an axenic medium have a slower growth rate of about 8-12 hours for doubling time.[5] It is often desirable to grow *Dictyostelium* cells as a pure culture in the absence of bacteria, however wild--type strains are unable to grow in axenic conditions. Cells grown in axenic mediums have slower growth rates, with a doubling time of roughly 8-12 hours depending on temperature, medium and the presence of selective drugs.[6] Growth can also vary significantly between different cell lines. Axenic-type strains can easily be transferred from bacterial growth conditions to axenic culture.[4]

**Time dependent plate analysis of strains**

All plated *D. discoideum* strains should be allowed to replicate for 12 hours to allow sufficient time for axenic-type strain growth to develop.[7] Plates should be viewed under a microscope to make initial observations and analyzed with ImageJ-plugin "ColonyArea". This tool would be used to quantitatively analyze area of Dictyostelium plaques in all the plates. Area of plaques would give an idea as to how fast or slow different species of Dictyostelium grow in presence of P. aeruginosa. The program works by processing data obtained through pictures of various dishes and separating organisms and formations based on color, texture, and size and eventually quantifying the data. The program does not count the total number of colonies, instead it determines the total percentage of area covered by the colonies and measuring the cell density of the area.[8] The experiment should be replicated multiple times to insure accuracy, and the results should be standardized by graphing ImageJ data with respect to time.

# Results
___________________________________________________________________________

The results of this experiment should support the hypothesis, that wild-type strain of *D. discoideum* was resistant or even less susceptible to infection caused by *P. aeruginosa*. However, axenic-type strain *D. discoideum* should be infected by the bacteria since it has remained isolated from other microbiota and has not developed bacterial resistivity mechanisms. New research shows that *D. discoideum* can grow symbiotic bacteria for nutrition by passing them along through generations and that these symbiotic bacteria can also protect the amoeba from environmental toxins.[9] There may also be membrane bound proteins that mediate amoeba defensive behaviors and play an essential role in bacterial resistivity.[17] If wild-type *D. discoideum* is able to resist the pathogenicity of *P. aeruginosa*, it will be evident by plaques formed on the plates where the amoeba feed on the bacteria. There should be no evidence of plaque formed on the axenic plates, since the bacteria will infects the amoeba and kill it. Based on these assumptions it is possible to analyze the plates using ImageJ software and quantify any differences between the two organisms by plaque formation. It is also key that all plates are analyzed at consistent time intervals to insure that plaque formation is not skewed by a factor of time. **FIG. 3** depicts bacterial plates being analyzed based on intensity using ImageJ-plugin "ColonyArea". Growth of *D. discoideum* plaques will be measured using this technique and the results will be quantified and compared between the different strains used.[19] Axenic-type strain *D. discoideum* is commonly used in research since it is cultivated over a period of time without the association to other microbes; hence, there is less error caused by extrinsic variables. Wild-type strains of *D. discoideum* associate with a multitude of organisms, therefore its behavior differs from axenic-type strains. Temperature and water accessibility may also play a role in *D. discoideum* resistance since they influence the overall development. Wild-type strain of *D. discoideum* obtained from a non-native prairie region, of Ben Geren Park, will be exposed to greater amounts of environmental stresses, such as more direct sunlight and chemicals, via resulting in the evolution of mechanisms to cope with these factors. However, it is expected that wild-type native prairie *D. discoideum* will only be influenced by microorganisms present in the soil. Overall, it is expected that both wild-type strains of *D. discoideum* strains will be more resistant to *P. aeruginosa*, while the axenic-type strain will be most susceptible to infection.

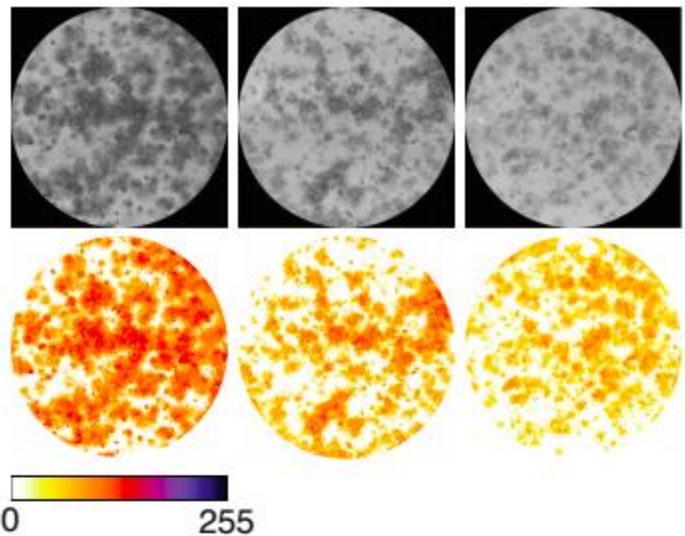

**FIG. 3.** Example of 3 plates being analyzed via intensity using ImageJ software[19]

Research has shown that the *Dictyostelium* model is predictive for *P. aeruginosa* virulence in a rat model.[2] It is expected that the results of pathogenicity experiments obtained using *D. discoideum* will be similar to experiments on rat (mammalian model), thus *Dictyostelium* is a simple non-mammalian system to study pathogenicity.

## Discussion
___________________________________________________________________

*P. aeruginosa* is an opportunistic bacterial pathogen that causes nosocomial infections in patients with compromised immune systems, such as urinary tract infections in pregnant women, cystic fibrosis, and cancer patients.[2] In this study, the wild-type *D. discoideum* is to be extracted from two locations. The first soil sample would come from a undisturbed prairie that contained limited pollutants. The second soil sample would come from a high traffic and well visited area that had been previously treated with chemicals. The overall purpose of the project is to confirm that environmental differences alter specific wild-type strains of *D. discoideum* that leads to adaptation and resistance to *P. aeruginosa*. Research studies have found that the slugs of *Dictyostelium* have small specialized cells that protect them.[16] Through generations of these cells, there may be developed capabilities that allow for protection against bacteria similar to *P. aeruginosa*. Based on this research, the wild-type strain of *D. discoideum,* collected from chemically treated soil, is expected to be the most resistant sample since it may possess specialized protective cells.[16] The Center for Disease Control states that the *P. aeruginosa* pathogen is one of the most common causes for bacterial infections, however, the effectiveness of treatment towards *P. aeruginosa* is decreasing since its becoming more resistant to antibiotics.[14] A project that may arise from this experimental proposal is identifying the species of *Dictyostelium* obtained from Ben Geren Park and comparing them to species of *Dictyostelium* known to the state of Arkansas. In a study that analyzed Dictyostelid cellular slime molds in Arkansas; 167 samples were collected and 2,082 individual clones representing 13 different species were identified, plus one form that could not be assigned to any described species.[18] Performing such an experiment may open new avenues of research and could lead to the discovery of new *Dictyostelium* clones. All in all, this research could be utilized in studies focused on studying antibiotic resistance in humans and may aid in the identification of other virulent pathways that bring about infection.

## Acknowledgements
___

Our sincere gratitude to Dr. Sandhya Baviskar for her guidance and encouragement in carrying out this biomedical research. A special thanks to other faculty members who rendered their help and knowledge. Research opportunity made possible through participation in Dr. Jeff Shaver's Undergraduate Biomedical Research course at the University of Arkansas Fort Smith.

## References
___